\documentclass[conference,letterpaper,10pt]{IEEEtran}
\usepackage{cite}
\usepackage{amsmath,amssymb}
\usepackage{graphicx}
\usepackage{flafter}
\usepackage{stfloats}
\usepackage{booktabs}
\usepackage{array}
\usepackage{microtype}
\usepackage{needspace}
\usepackage{url}
\usepackage{threeparttable}
\usepackage[hidelinks]{hyperref}
\graphicspath{{figures/}}
\title{SimSkip: StrongARM-Comparator Similarity Sensing with SAR-Skipping-Aware Training for Analog Compute-in-Memory}
\author{\IEEEauthorblockN{Yufei Liu, Junjie Wang\textsuperscript{*}, Shuang Liu, Yang Wang, and Yang Liu}
\IEEEauthorblockA{University of Electronic Science and Technology of China\\
Chengdu,
China\\
\textsuperscript{*}Corresponding author: Junjie Wang(wangjunjie@uestc.edu.cn)}}

\begin{document}
\maketitle

\clubpenalty=10000
\widowpenalty=10000
\setlength{\abovedisplayskip}{4pt plus 2pt minus 2pt}
\setlength{\belowdisplayskip}{4pt plus 2pt minus 2pt}
\raggedbottom

\begin{abstract}
Analog-to-digital conversion dominates the energy and area of analog compute-in-memory (CIM) accelerators.
This paper presents SimSkip,
a successive-approximation-register (SAR) skipping scheme that combines a lightweight StrongARM-comparator similarity detector with SAR-skipping-aware training.
When two held voltages are similar,
the upper bits of the preceding code are reused to skip SAR comparisons.
Detector-induced disturbance extracted from 55-nm CMOS simulations and bit-exact prefix saturation are embedded in the forward pass,
and the network learns to tolerate them through quantization-aware training and knowledge distillation.
On an ISAAC-based W4A4 WRN-28-10,
SimSkip achieves 77.23\% CIFAR-100 top-1 accuracy (mean of three seeds) versus 78.4\% for the baseline,
while reducing SAR comparisons by 48.2\% across three hardware-modeled layers.
A first-order model estimates a 27.2\% net ADC energy reduction after detector overhead.
\end{abstract}


\section{Introduction}
Compute-in-memory (CIM) performs matrix-vector multiplication inside memory arrays,
alleviating the data-movement bottleneck of von~Neumann architectures for edge AI inference~\cite{verma2019imc,sebastian2020memory}.
In analog CIM,
however,
analog-to-digital conversion is a major bottleneck: in ISAAC,
ADCs account for 58\% of the tile power and 31\% of the tile area~\cite{shafiee2016isaac},
motivating several ADC-reduction techniques~\cite{andrulis2023raella,fu2022cim,jo2025similarity}.
Neighboring CIM outputs often exhibit value similarity.
Jo \textit{et al.}~\cite{jo2025similarity} exploited this locality with auxiliary comparators that bypass selected SAR decisions.

We propose SimSkip,
which uses StrongARM comparators to detect similarity between two analog CIM results and guide SAR skipping in the next conversion.
A StrongARM comparator resolves its input difference through positive feedback within a fixed clock window,
enabling low-energy threshold sensing;
double-tail, dynamically biased, and rail-to-rail complementary-input variants further improve speed and energy efficiency~\cite{schinkel2007doubletail,bindra2017dynamic,alqadasi2020rail}.
However,
aggressive skipping makes the readout inexact.
Comparator kickback disturbs the held voltages,
and when two similar inputs lie in different prefix-defined subranges,
the reused prefix confines the SAR search to the wrong subrange and the output saturates at the nearest subrange limit (prefix saturation).
Rather than suppressing these errors with extra circuitry,
SimSkip models both as deterministic functions of the held voltages,
embeds them in the forward pass,
and trains the network end-to-end to absorb them,
which keeps the detector lightweight and enables aggressive skip lengths.
Our contributions are:
(1)~an end-to-end SAR-skipping-aware training method,
built on quantization-aware training (QAT) and knowledge distillation,
that executes a circuit-extracted readout model;
(2)~a two-sided similarity detector in which two complementary-input StrongARM comparators separately evaluate the upper and lower thresholds and provide the characterized disturbance used in training;
and (3)~SimSkip,
a similarity-guided SAR-skipping scheme built on the ISAAC architecture.

\section{Proposed SimSkip Scheme}

\subsection{Similarity-Guided SAR Bit Skipping}
\label{sec:skipping}
ISAAC~\cite{shafiee2016isaac} is a representative analog CIM accelerator whose tile organization has been adopted by many later designs.
In each in-situ multiply-accumulate unit,
input activations drive the crossbar word lines,
each bit line accumulates the weighted currents of a column,
and sample-and-hold (S\&H) circuits hold the column voltages for a time-multiplexed SAR ADC.
SimSkip adds a similarity detector that lets this shared ADC skip redundant comparisons.

\begin{figure}[!htbp]
    \centering
    \includegraphics[width=0.72\columnwidth]{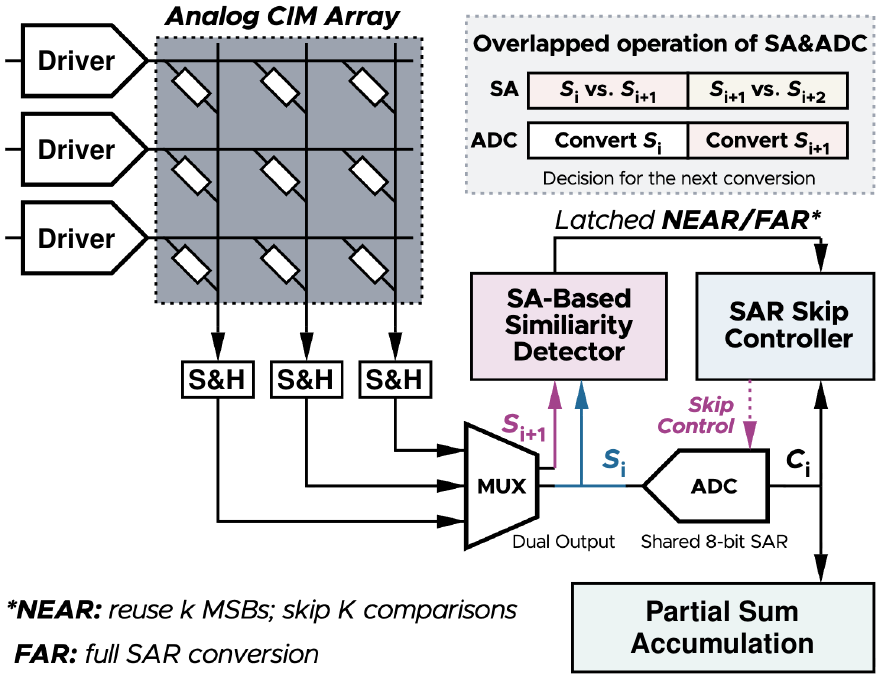}
    \caption{Proposed SimSkip readout architecture,
    with similarity detection overlapped with SAR conversion.}
    \label{fig:architecture}
\end{figure}
As shown in Fig.~\ref{fig:architecture},
$S_i$ denotes the held voltage of the $i$th bit line,
and the shared ADC is a single-ended 8-bit SAR ADC.
A dual-output multiplexer (MUX) routes $S_i$ and $S_{i+1}$ to the ADC and the detector simultaneously.
While the ADC converts $S_i$ into the code $c_i$,
the detector classifies the pair as NEAR if $|S_{i+1}-S_i|\leq V_T$ and as FAR otherwise,
where $V_T$ is the similarity threshold.
For a NEAR pair,
the controller presets the upper $k$ bits ($k$-bit MSB prefix) of the conversion of $S_{i+1}$ to those of $c_i$ and skips $k$ comparisons;
a FAR pair triggers a full conversion.

Reusing a $k$-bit prefix in a $B$-bit ADC restricts the remaining conversion to a residual range $V_{\mathrm{R}}(k)=2^{B-k}V_{\mathrm{LSB}}=V_{\mathrm{FS}}/2^k$,
where $V_{\mathrm{FS}}$ is the full-scale range and $V_{\mathrm{LSB}}$ is the quantization step.
For $V_{\mathrm{FS}}=1.2$~V and $B=8$,
$V_{\mathrm{LSB}}=4.6875$~mV,
$V_{\mathrm{R}}(4)=75$~mV,
and $V_{\mathrm{R}}(6)=18.75$~mV.
The design-time pair $(V_T,k)$ sets the trade-off:
a larger $V_T$ admits more NEAR pairs and a larger $k$ skips more comparisons per NEAR pair,
but both make a NEAR pair more likely to span two subranges,
causing the prefix saturation modeled in Section~\ref{sec:hardware-aware}.

\subsection{Two-Sided Similarity Detector}
\label{sec:detector}
\begin{figure}[!tb]
    \centering
    \includegraphics[width=0.74\columnwidth]{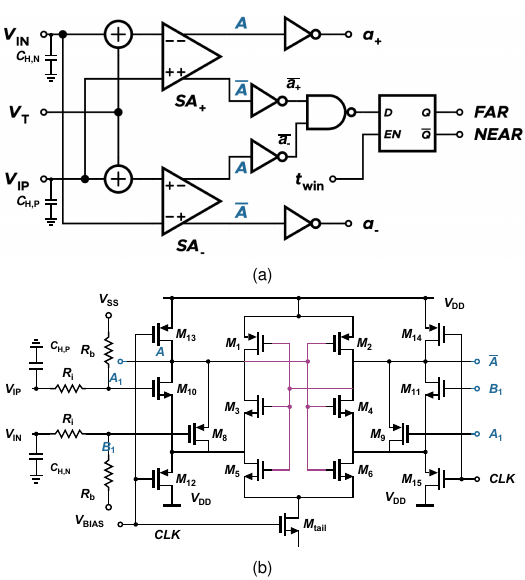}
    \caption{Two-sided similarity detector.
    (a) Two-sided comparison and decision circuit.
    (b) Rail-to-rail complementary-input StrongARM comparator and threshold-offset generation circuit.}
    \label{fig:sa_detector}
\end{figure}

As shown in Fig.~\ref{fig:sa_detector},
the detector consists of two rail-to-rail complementary-input StrongARM comparators adapted from~\cite{alqadasi2020rail},
a threshold-offset generation circuit,
and CMOS decision and latch logic.
The positive input $V_{\mathrm{IP}}$ connects to the S\&H node of $S_{i+1}$ and the negative input $V_{\mathrm{IN}}$ to that of $S_i$.
CMP$_+$ asserts $a_+$ when $V_{\mathrm{IP}}-V_{\mathrm{IN}}>V_T$,
and CMP$_-$ asserts $a_-$ when $V_{\mathrm{IN}}-V_{\mathrm{IP}}>V_T$,
defining the similarity interval $[-V_T,+V_T]$.
In each comparator,
a fixed-ratio resistive divider derives the threshold offset from the bias voltage,
and a series capacitor couples it into the input path.
A NAND gate combines $\overline{a_+}$ and $\overline{a_-}$ into $\mathrm{FAR}=a_+\lor a_-$,
and a D latch stores FAR and its complement,
NEAR,
for the SAR controller (Fig.~\ref{fig:sa_regeneration}).
We implement a fine detector,
D19 ($V_T=19$~mV),
and a coarse detector,
D38 ($V_T=38$~mV).

\subsection{Nonideality Modeling}
\label{sec:hardware-aware}
The readout model includes detector-induced disturbance and prefix saturation,
with the disturbance extracted from circuit simulations at TT, 1.2~V, and 27$^\circ$C.

\begin{figure}[!tb]
\centering
\includegraphics[width=0.88\columnwidth]{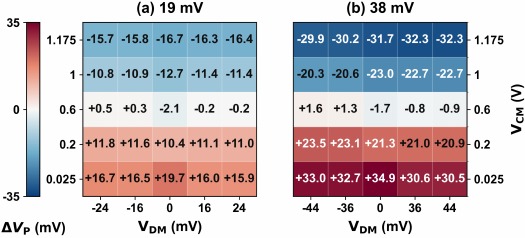}
\caption{Detector-induced held-voltage shifts $\Delta V_P$ for (a) D19 and (b) D38 at TT, 1.2~V, and 27$^\circ$C.}
\label{fig:disturbance_map}
\end{figure}

Kickback from capacitive coupling of regenerating internal nodes and charge changes in the input transistors~\cite{figueiredo2006kickback,lei2013systematic},
together with loading,
shifts the held voltages at $V_{\mathrm{IP}}$ and $V_{\mathrm{IN}}$ by $\Delta V_P$ and $\Delta V_N$,
respectively.
With the nominal circuit and timing conditions fixed,
SimSkip models each shift as a deterministic function of the held voltages,
extracted as the difference between node voltages simulated with and without the detector at the same instant.
With $V_{\mathrm{hold}}$ denoting the undisturbed held voltage of $S_{i+1}$,
the single-ended ADC sees
\begin{equation}
    V_{\mathrm{ADC}}=V_{\mathrm{hold}}+\Delta V_P,\qquad
\Delta V_P=f_{\mathrm{SA}}(V_{\mathrm{CM}},V_{\mathrm{DM}}),
\label{eq:disturb}
\end{equation}
where $V_{\mathrm{CM}}=(V_{\mathrm{IP}}+V_{\mathrm{IN}})/2$,
$V_{\mathrm{DM}}=V_{\mathrm{IP}}-V_{\mathrm{IN}}$,
and $f_{\mathrm{SA}}(\cdot)$ is a two-dimensional lookup table that interpolates the shifts simulated at the nonuniform input points in Fig.~\ref{fig:disturbance_map}.
The disturbed voltage and its code $c_{i+1}$ serve as the reference for the next comparison and prefix reuse.
The shifts are sampled at 985~ps,
after the decision window closes at 900~ps;
across the 25 input pairs per class,
the maximum $|\Delta V_P|$ is 19.69~mV for D19 and 34.88~mV for D38.

If a NEAR pair has different $k$-bit prefixes,
the reused prefix confines the SAR search to the wrong subrange,
and the output saturates at the subrange limit nearest to $S_{i+1}$
(e.g., $0111$ instead of $1000$ for $k=1$ in a 4-bit ADC),
so the error is bounded by the input difference ($\leq V_T$).
SimSkip reproduces this bit-exactly by executing the SAR search with the reused prefix.

\subsection{SAR-Skipping-Aware Training}
\label{sec:training}
\begin{figure}[!tb]
    \centering
    \includegraphics[width=.65\columnwidth]{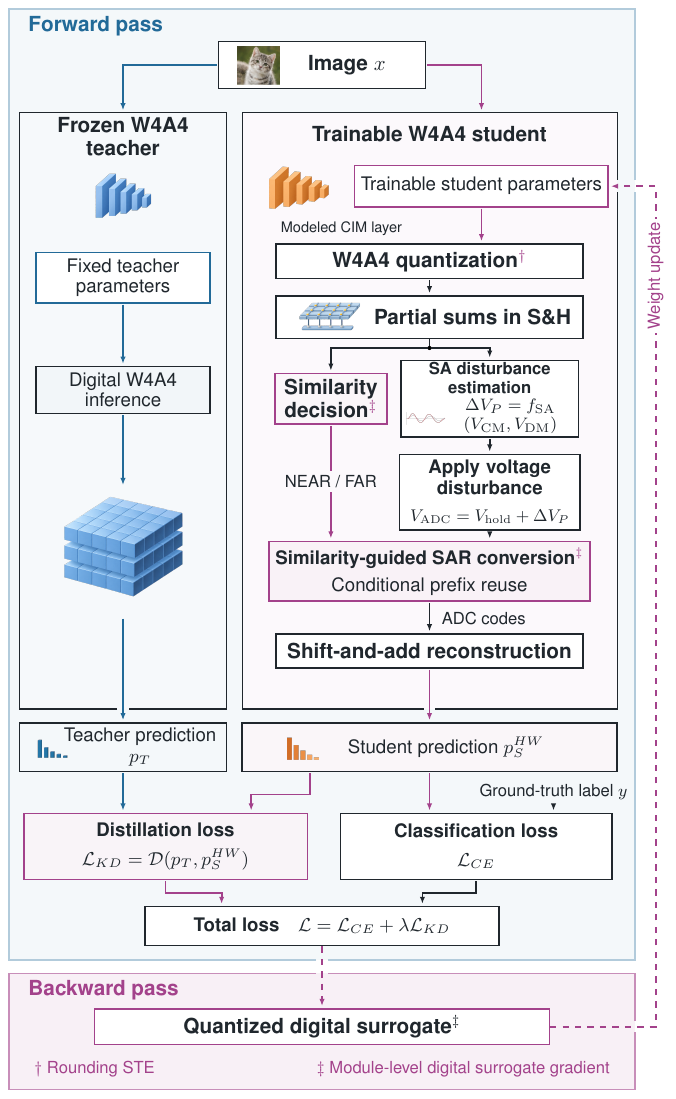}
    \caption{SAR-skipping-aware distillation with a frozen W4A4 teacher and a trainable hardware-modeled W4A4 student.}
    \label{fig:hwa_training}
\end{figure}

SAR-skipping-aware training is a form of hardware-aware training (HAT)~\cite{rasch2023hardware} built on QAT and uses two W4A4 networks (Fig.~\ref{fig:hwa_training}).
Standard QAT of the FP32 baseline with an ideal readout produces the W4A4 teacher,
which is then frozen.
The student is initialized from the teacher and replaces the ADC readout of the hardware-modeled layers with the SimSkip model of Sections~\ref{sec:skipping}--\ref{sec:hardware-aware}.
For the same input sample,
the student minimizes
\begin{equation}
    \mathcal{L}=\mathcal{L}_{\mathrm{CE}}+\lambda\mathcal{L}_{\mathrm{KD}},\quad
    \mathcal{L}_{\mathrm{KD}}=D_{\mathrm{KL}}\!\left(p_T(\tau)\,\|\,p_S^{\mathrm{HW}}(\tau)\right),
    \label{eq:training_loss}
\end{equation}
where $\mathcal{L}_{\mathrm{CE}}$ is the cross-entropy with the ground truth,
$\mathcal{L}_{\mathrm{KD}}$ is the distillation loss~\cite{hinton2015distilling},
and $p_T(\tau)$ and $p_S^{\mathrm{HW}}(\tau)$ are the teacher and hardware-modeled student softmax probabilities at temperature $\tau$,
which softens the distributions so that the student learns the teacher's relative class probabilities.
We use $\tau=2$ and $\lambda=2$.
Weights and activations use 4-bit per-layer quantization with an identity straight-through estimator (STE)~\cite{bengio2013estimating}.
Each hardware-modeled layer executes the full readout model (disturbance, NEAR/FAR classification, prefix reuse, ADC encoding, and reconstruction) in the forward pass,
and the backward pass uses the gradient of the ideal quantized layer (a module-level STE).
$V_T$ and $k$ stay fixed,
and only the trained student is deployed.

\section{Evaluation and Results}
\subsection{Detector Characterization in 55-nm CMOS}
\label{sec:char}
The detector is simulated at the transistor level in 55-nm CMOS,
with each input driven by an ideal voltage source and held on a dedicated 50-fF capacitor.
An \textit{event} is one complete detector operation (reset, regeneration, and latching) for a given $(V_{\mathrm{CM}},V_{\mathrm{DM}})$ pair.
An event passes when both comparator decisions and the latched NEAR/FAR output are correct,
the reset completes,
and the event energy stays within 80~fJ.
The event period is the interval between successive event starts,
and the decision availability time is the delay from the event start to a valid latched NEAR/FAR output.

Table~\ref{tab:detector_char} summarizes the results at TT, 1.2~V, and 27$^\circ$C with a 2.5-ns event period;
the event counts include both NEAR and FAR inputs.
All events pass,
with decision availability within 760~ps.
The reported energy is the largest per-event total drawn from the comparator,
decision/storage,
bias,
and CLK/WINDOW supply ports,
excluding the analog sources and the internal losses of the bias and timing generators.
The 57.10/61.31~fJ per event for D19/D38,
which covers two comparators and the decision logic,
is comparable with 65-nm dynamic comparators reporting 15.2~fJ/comparison (post-layout) in~\cite{alqadasi2020rail} and 30~fJ/comparison in~\cite{bindra2017dynamic}.

\begin{table}[!tb]
\centering
\begin{threeparttable}
\caption{Detector Characterization at TT, 1.2~V, and 27$^\circ$C}
\label{tab:detector_char}
\footnotesize
\setlength{\tabcolsep}{4pt}
\begin{tabular}{lcc}
\toprule
\textbf{Parameter} & \textbf{D19} & \textbf{D38} \\
\midrule
Threshold $V_T$ & 19 mV & 38 mV \\
Valid $V_{\mathrm{CM}}$ sweep (43 points) & 0.012--1.188 V & 0.022--1.178 V \\
Tested $V_{\mathrm{DM}}$ (mV) & $0,\pm16,\pm24$ & $0,\pm36,\pm44$ \\
Passed / valid events\tnote{a} & 390/390 & 342/342 \\
Decision availability time & $\leq$760 ps & $\leq$760 ps \\
Max. energy per event & 57.10 fJ & 61.31 fJ \\
\bottomrule
\end{tabular}
\begin{tablenotes}[flushleft]
\scriptsize
\item[a] $43\times5=215$ nominal combinations per drive condition;
those with $V_{\mathrm{IP}}$ or $V_{\mathrm{IN}}$ outside 0--1.2~V are excluded,
leaving 195 (D19) and 171 (D38),
summed over two drive conditions.
\end{tablenotes}
\end{threeparttable}
\end{table}

\begin{figure}[!tb]
    \centering
    \includegraphics[width=0.76\columnwidth]{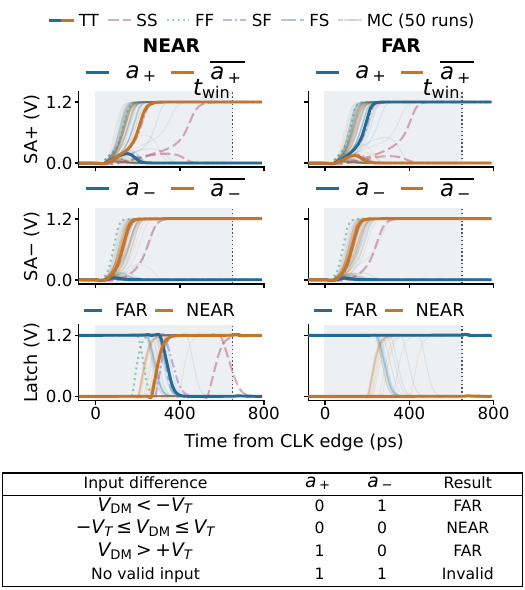}
    \caption{D38 comparator and latch waveforms across five process corners and 50 mismatch Monte Carlo runs.}
    \label{fig:sa_regeneration}
\end{figure}

Table~\ref{tab:pvt_axes} reports the PVT characterization,
sweeping each axis separately with 25 events (15 NEAR, 10 FAR) per class at each point.
From the event start,
the decision window closes at 0.900~ns and outputs are checked at 0.985~ns within a 2.5-ns period
(1.850, 1.985, and 5~ns in the voltage and temperature tests).
All events pass across five process corners,
0.85--1.2~V,
and $-40$ to 125$^\circ$C.
Fig.~\ref{fig:sa_regeneration} shows D38 waveforms at 1.2~V and 27$^\circ$C
for the process corners ($V_{\mathrm{CM}}=0.6$~V)
and for 50 Monte Carlo runs at TT with mismatch in the comparator pair, NAND gate, and latch ($V_{\mathrm{CM}}=25$~mV).

\begin{table}[!tb]
\centering
\caption{Separate-Axis PVT Characterization}
\label{tab:pvt_axes}
\footnotesize
\setlength{\tabcolsep}{2pt}
\renewcommand{\arraystretch}{0.95}
\begin{tabular}{@{}>{\raggedright\arraybackslash}p{\dimexpr.15\columnwidth-3pt\relax}>{\raggedright\arraybackslash}p{\dimexpr.24\columnwidth-3pt\relax}>{\raggedright\arraybackslash}p{\dimexpr.34\columnwidth-3pt\relax}>{\raggedright\arraybackslash}p{\dimexpr.27\columnwidth-3pt\relax}@{}}
\toprule
\textbf{Sweep} & \textbf{Fixed conditions} & \textbf{Test conditions} & \textbf{Passed events} \\
\midrule
Process & 1.2~V, 27$^\circ$C & TT, SS, FF, FS, SF & 250/250 \\
Voltage & TT, 27$^\circ$C & 0.85--1.20~V (6 points) & 300/300 \\
& TT, 27$^\circ$C & 0.825~V & 23/25 per class \\
& TT, 27$^\circ$C & 0.800~V & 12/25 (D19);\newline 13/25 (D38) \\
Temp. & TT, 1.2~V & $-40$--125$^\circ$C (7 points) & 350/350 \\
\bottomrule
\multicolumn{4}{@{}p{\columnwidth}@{}}{\scriptsize $V_{\mathrm{CM}}=\{0.025,0.2,0.6,1.0,1.175\}$~V at 1.2~V; $V_{\mathrm{DM}}$: D19 $\{0,\pm16,\pm24\}$~mV, D38 $\{0,\pm36,\pm44\}$~mV. Voltage sweep: inputs and $V_T$ scale by $V_{\mathrm{DD}}/(1.2\,\mathrm{V})$.}
\end{tabular}
\end{table}

\subsection{End-to-End Accuracy--SAR Trade-off}
We evaluate SimSkip on CIFAR-100~\cite{krizhevsky2009learning} using the open-source ISAAC model of CiMLoop~\cite{andrulis2024cimloop} with W4A4 quantization.
WRN-28-10~\cite{zagoruyko2016wide} is the primary case study,
and ResNet20~\cite{he2016deep} and the vision Transformer DeiT~\cite{touvron2021training} assess cross-model applicability.
The three-layer mapping assigns the readout model to the second convolution of the last block in each WRN stage;
\texttt{layer1.0.conv1}, \texttt{layer2.1.conv2}, and \texttt{layer3.2.conv2} in ResNet20;
and the second MLP linear layer of blocks 1, 6, and 12 in DeiT.
The SAR-comparison reduction $R_{\mathrm{cmp}}$ is computed from the total comparison counts over the mapped layers and all evaluation samples,
weighting each layer by its number of conversions.

The three-layer WRN student is fine-tuned from the W4A4 teacher with (\ref{eq:training_loss}) for 20 epochs using AdamW,
with cosine learning-rate decay from $10^{-5}$ to $10^{-6}$.
The batch-normalization (BN) statistics are then reset,
re-estimated under hardware errors by cumulative averaging over 2000 unaugmented training samples (batch size 32) without weight updates,
and frozen.

\begin{figure}[!tb]
    \centering
    \includegraphics[width=0.92\columnwidth]{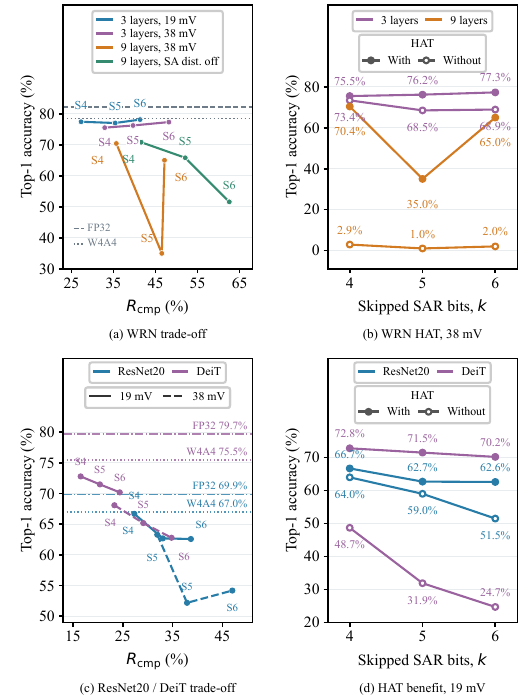}
    \caption{End-to-end accuracy and SAR-skipping results: (a) WRN trade-offs for three- and nine-layer mappings, including the disturbance-disabled diagnostic; (b) WRN HAT benefit with D38; (c) ResNet20 and DeiT trade-offs; (d) their HAT benefit with D19.}
    \label{fig:end_to_end_results}
\end{figure}

The FP32 and W4A4 WRN baselines reach 82.1\% and 78.4\% top-1 accuracy.
Fig.~\ref{fig:end_to_end_results}(a)--(b) evaluates the three-layer mapping with D19/D38 and $k=4$, 5, and 6 (S4/S5/S6).
The plotted D38/S6 run, also used in Table~\ref{tab:energy}, reaches 77.3\% with a 48.19\% comparison reduction;
three seeds from the same W4A4 checkpoint and data splits give $(77.23\pm0.12)\%$ and $(48.24\pm0.07)\%$.
When WRN's second residual group (nine convolutions, including the projection) is mapped with D38,
30 HAT epochs yield 70.4/35.0/65.0\% for S4/S5/S6 on 1,000 test images,
versus 2.9/1.0/2.0\% for the untrained controls with BN calibration (Fig.~\ref{fig:end_to_end_results}(a)--(b)).
With disturbance disabled at evaluation (S5-trained weights, BN recalibrated per skip length),
accuracy falls monotonically to 70.8/65.8/51.6\% on 500 validation images,
so the non-monotonic S5 result stems from disturbance rather than prefix saturation.

Under identical inference errors,
training also improves ResNet20 and DeiT over the untrained controls (Fig.~\ref{fig:end_to_end_results}(c)--(d)):
both CNNs (with BN calibration) approach their W4A4 baselines at D19/S4,
and DeiT rises from 48.7\% to 72.8\% at D19/S4 and from 24.7\% to 70.2\% at D19/S6.
ResNet20 alone loses pronounced activation sparsity when trained with disturbance;
training with prefix saturation alone preserves it,
and thus zero-gating opportunities,
at a small accuracy penalty.

\subsection{Energy Analysis}
The analysis covers ADC conversion and similarity detection,
excluding routing, control, and other peripheral energy.
The Accelergy ADC model in our CiMLoop-based ISAAC evaluation~\cite{andrulis2024cimloop} gives a full-conversion energy of $E_{\mathrm{ADC}}=1.929$~pJ for the 8-bit SAR ADC.
Following~\cite{xie2021skipping},
we split it as $E_{\mathrm{ADC}}=E_{\mathrm{fix}}+E_{\mathrm{comp}}$,
where $E_{\mathrm{comp}}$ is proportional to the number of executed comparisons.
Calibrating with Jo \textit{et al.}~\cite{jo2025similarity},
who report a 31.5\% ADC-power reduction ($r_{\mathrm{ref}}=0.315$) when three of six comparisons are bypassed ($f_{\mathrm{skip}}=3/6$),
gives $E_{\mathrm{comp}}/E_{\mathrm{ADC}}=r_{\mathrm{ref}}/f_{\mathrm{skip}}\approx0.63$ and $E_{\mathrm{fix}}/E_{\mathrm{ADC}}\approx0.37$.
For a comparison reduction $R_{\mathrm{cmp}}$ and one detector event of energy $E_{\mathrm{det}}$ per conversion,
the gross and net ADC energy reductions are
\begin{equation}
R_{\mathrm{gross}}\approx0.63R_{\mathrm{cmp}},\qquad
R_{\mathrm{net}}=R_{\mathrm{gross}}-E_{\mathrm{det}}/E_{\mathrm{ADC}}.
\end{equation}
At D38/S6 (Table~\ref{tab:energy}),
the gross saving is $1.929~\mathrm{pJ}\times30.36\%\approx0.586$~pJ per conversion;
subtracting the 0.061-pJ detector energy leaves about 0.524~pJ,
or 27.18\%.

\begin{table}[!tb]
\centering
\caption{ADC Energy Reduction Estimates}
\label{tab:energy}
\footnotesize
\setlength{\tabcolsep}{3pt}
\begin{tabular}{@{}lrrr@{}}
\toprule
\textbf{Operating point} & \textbf{SAR comp.} & \textbf{Gross} & \textbf{Net} \\
\midrule
D19 / S4 (this work) & 27.30\% & 17.20\% & 14.24\% \\
D38 / S6 (this work) & 48.19\% & 30.36\% & 27.18\% \\
Jo \textit{et al.}~\cite{jo2025similarity} (65 nm) & $\sim$24.5\%$^\dagger$ & 31.5\%$^\ddagger$ & 5.1\%$^\ddagger$ \\
\bottomrule
\multicolumn{4}{@{}p{\columnwidth}@{}}{\scriptsize $^\dagger$$49\%\times3/6$; auxiliary comparisons excluded.
$^\ddagger$Reported ADC-power and workload-level SA-ADC energy reductions; accounting scopes differ, and no technology normalization is applied.} \\
\end{tabular}
\end{table}

\section{Conclusion}
SimSkip uses end-to-end training to absorb readout errors,
enabling lightweight StrongARM-comparator similarity detection and aggressive SAR-prefix reuse in analog CIM.
Trained with a circuit-extracted forward model of detector disturbance and prefix saturation,
all three networks improve accuracy over untrained controls,
while SAR comparisons in the hardware-modeled layers drop by up to 48.24\%,
giving an estimated 27.18\% net ADC energy reduction after detector overhead.
Future work will add random perturbations for device mismatch and clock-phase noise to the training model.

\clearpage
\bibliographystyle{IEEEtran}
\bibliography{references}

\end{document}